\documentclass[a4paper,12pt]{article}
\PassOptionsToPackage{table,dvipsnames}{xcolor}
\usepackage{jheppub} % for details on the use of the package, please
\usepackage{dcolumn}% Align table columns on decimal point
\usepackage{bm}% bold math
\usepackage{physics}
\usepackage{cancel}

\usepackage[table]{xcolor}
\usepackage{todonotes}

\definecolor{darkpink}{RGB}{153,0,76}

\usepackage[most]{tcolorbox}
\newcommand\eqbox[1]{\tcbhighmath{#1}}
\tcbset{highlight math style={left=0mm,right=0mm,top=0mm,bottom=0mm}}

\presetkeys{todonotes}{disable}{} %MHV to turn-off  todo

\DeclareMathAlphabet{\mathpzc}{OT1}{pzc}{m}{it} %New Font
\newcommand{\be}{\begin{equation}}
	\newcommand{\bea}{\begin{eqnarray}}
		\newcommand{\eea}{\end{eqnarray}}
	\newcommand{\ba}{\begin{align}}
		\newcommand{\ea}{\end{align}}
	\newcommand{\ee}{\end{equation}}

\usepackage{tikz}
\newcommand{\Plus}{\mathord{\begin{tikzpicture}[baseline=0ex, line width=.8, scale=0.07]
\draw (1,0) -- (1,2);
\draw (0,1) -- (2,1);
\end{tikzpicture}}}
\newcommand{\Minus}{\mathord{\begin{tikzpicture}[baseline=0ex, line width=.8, scale=0.07]
\draw (0,1) -- (2,1);
\end{tikzpicture}}}

\newcommand{\tL}{l}
\newcommand{\F}{\mathpzc{F}}
\newcommand{\G}{\mathpzc{G}}
\newcommand{\EW}{E_{W}}
\newcommand{\tEW}{\tilde{E}_W}
\newcommand{\Q}{\mathcal{Q}}
\newcommand{\A}{{\mathpzc{A}}}
\newcommand{\EWa}{\EW^{\;\Minus}}
\newcommand{\EWb}{\EW^{\;\Plus}}
\newcommand{\tEWa}{\tEW^{\;\Minus}}
\newcommand{\tEWb}{\tEW^{\;\Plus}}
\newcommand{\Qa}{\Q_{\;\Minus}}
\newcommand{\Qb}{\Q_{\;\Plus}}
\newcommand{\Aa}{\A_{\;\Minus}}
\newcommand{\Ab}{\A_{\;\Plus}}
\newcommand{\rh}{r_0}
\newcommand{\um}{{\bmu_{{\rm m}}}}
\newcommand{\uws}{\bmu_{\star}}
\DeclareSymbolFont{anttfont}{OML}{antt}{m}{it}
\DeclareMathSymbol{\CA}{\mathalpha}{anttfont}{`c}
\DeclareMathSymbol{\bmu}{\mathalpha}{anttfont}{`u}
\DeclareMathAlphabet{\mathcalligra}{T1}{calligra}{m}{n}%
\makeatletter  %%%MHV to remove JHEP Header
\gdef\@fpheader{}  %%%MHV to remove JHEP Header
\makeatother %%%MHV to remove JHEP Header

\begin{document}
%\listoftodos
\newpage

\preprint{IPM/P-2023/nnn}

\title{\Large\centering {Entanglement Wedge Cross Section \\Growth During Thermalization }}
%\title{\Large \centering {Reflected Entropy Growth During Thermalization }}
%\title{\centering {Thermalization and Growth of \\Reflected Entropy }}

\author[\star,\dagger]{Komeil Babaei Velni}
\author[\star,\dagger]{M. Reza Mohammadi Mozaffar}
\author[\natural,\dagger]{M.H.~Vahidinia}

\affiliation[\star]{ Department of Physics, University of Guilan,	P.O. Box 41335-1914, Rasht, Iran}
\affiliation[\natural]{Department of Physics, Institute for Advanced Studies in Basic Sciences (IASBS),	P.O. Box 45137-66731, Zanjan, Iran}
\affiliation[\dagger]{School of Physics, Institute for Research in Fundamental
Sciences (IPM), P.O.Box 19395-5531, Tehran, Iran}

\emailAdd{babaeivelni@guilan.ac.ir, mmohammadi@guilan.ac.ir, vahidinia@iasbs.ac.ir}

\abstract{Motivated by exploring the thermalization process in relativistic and non-relativistic holographic field theories after a non-local quench, we investigate some features in the time evolution of the entanglement wedge cross section (EWCS). This quantity is a possible holographic dual to some non-local information measures such as entanglement of purification. In particular, we focus on the time dependence of EWCS  during black hole formation in $D+2$ dimensional AdS spacetime as well as geometries with Lifshitz and hyperscaling violating exponents.  A combination of analytic and numerical results for large symmetric strip shaped boundary subregions shows that the scaling of EWCS at early times only depends on the Lifshitz exponent. In addition, this early growth regime is followed by a linear-growth regime whose velocity depends on the dimensions of spacetime, the Lifshitz exponent, and the hyperscaling parameter. This velocity is the same as the entanglement velocity and for nontrivial dynamical exponent depends on the temperature of  the final equilibrium state.}

\date{\today}% It is always \today, today,
%  but any date may be explicitly specified
\maketitle
%%%%%%%%%%%%%%%%%%%%%%%%%%%%%%%
\section{Introduction}\label{intro}
%%%%%%%%%%%%%%%%%%%%%%%%%%%%%%%
The gauge/gravity duality allows us to quantitatively study surprising new connections between quantum
information theory and quantum gravity in recent years, \textit{e.g.}, see reviews \cite{Rangamani:2016dms,Nishioka:2018khk,Casini:2022rlv}. In particular, it is now evident that certain geometric
quantities in the bulk geometry can be related to the information-theoretic measures of the boundary field theory. In this context, the entanglement entropy (EE) associated with a spatial boundary subregion $A$ is determined by the Ryu-Takayanagi (RT) formula \cite{Ryu:2006bv}
\begin{eqnarray}\label{RT}
S_A={\rm min}\frac{{\rm Area}(\Gamma_A)}{4G_N},
\end{eqnarray}
where $\Gamma_A$ is a bulk minimal hypersurface homologous to $A$, \textit{i.e.,} $\partial \Gamma_A=\partial A$. Moreover, the Hubeny-Rangamani-Takayanagi (HRT) proposal \cite{0705.0016} extends this prescription to time-dependent situations by considering extremal hypersurfaces with the same boundary condition. These proposals have stimulated a wide variety of research efforts investigating the properties of entanglement and information measures holographically. An interesting suggestion in this research program is that the entanglement of mixed states in the
boundary theory is encoded in a certain codimension-two bulk hypersurface which is called the entanglement wedge cross section (EWCS). Considering a spatial region $A$ in the boundary theory, the entanglement wedge is the bulk region corresponding to the reduced density matrix $\rho_A$ and whose boundary is $A\cup \Gamma_A$. In particular, when the boundary region is the union of two disjoint subregions $A_1$ and $A_2$ the boundary of the entanglement wedge is $A_1\cup A_2\cup\Gamma_{A_1\cup A_2}$. For small separations where the connected configuration is favored, the EWCS is defined to be the minimal cross sectional area of the entanglement wedge
and is given as follows \cite{Takayanagi:2017knl, Nguyen:2017yqw} (see \textit{e.g.} figure \ref{fig:regions} )
\begin{eqnarray}\label{eq:EwCS}
\EW(A_1, A_2)={\rm ext}\frac{{\rm Area}(\Sigma_{A_1\cup A_2})}{4G_N}.
\end{eqnarray}
On the other hand, for large separations, the disconnected configuration is favored and the EWCS vanishes. Considering the above definition, different proposals that make connections between the EWCS and boundary correlation measures can be recapped as follows \cite{Takayanagi:2017knl,Nguyen:2017yqw,Dutta:2019gen,Tamaoka:2018ned}
\begin{eqnarray*}
\EW(A_1, A_2)=E_P(A_1, A_2)=\frac{S_R(A_1, A_2)}{2}=S_O(A_1, A_2)-S(A_1 \cup A_2),
\end{eqnarray*}
where $E_P$, $S_R$, and $S_O$ are entanglement of purification, reflected entropy, and odd entropy respectively.\footnote{Recently in \cite{Hayden:2023yij} it was shown that $S_R$ is not monotonically decreasing under partial trace, and so in general is not a measure of physical correlations. However, it seems that it is a valid correlation measure for holographic states as entanglement wedge nesting suggests\cite{Wall:2012uf,Dutta:2019gen}.} One may note that although they are not necessarily equivalent for a generic state, it seems they are the same for holographic states.
Indeed, the EWCS may probe spacetime properties that are inaccessible from the perspective of holographic entanglement entropy (HEE). In particular, HEE is not a measure of quantum entanglement for a mixed state. This progress has motivated some interesting and extensive discussions of holographic correlation measures which have led to a remarkably rich and varied range of new insights in both holography and field theory, \textit{e.g.,}  \cite{Hirai:2018jwy,BabaeiVelni:2019pkw,Jokela:2019ebz,Umemoto:2019jlz,Akers:2019gcv,Amrahi:2020jqg,Chakrabortty:2020ptb,Saha:2021kwq,1810.00420,1907.06646,Jeong:2019xdr,2001.05501,Moosa:2020vcs,Boruch:2020wbe,Khoeini-Moghaddam:2020ymm,Basu:2022nds,Basak:2022cjs,1909.06790,Mollabashi:2020ifv,Berthiere:2020ihq,Bueno:2020fle,Sahraei:2021wqn,Camargo:2021aiq,Wen:2021qgx,Chowdhury:2021idy,Hayden:2021gno,Akers:2021pvd,Bueno:2020vnx,Camargo:2022mme,ChowdhuryRoy:2022dgo,Vasli:2022kfu}.

At the same time, there has been a great deal of interest in studying quantum quenches to understand whether and how quantum matter equilibrates. A prime arena for discussions of holographic quantum quench has been the Vaidya spacetime and this will also be the case in the present paper. This bulk geometry
describes the gravitational collapse of a thin shell of matter to form a black hole which is dual to the evolution of a far from equilibrium initial state to a steady state in the boundary theory. Such shock-wave geometries have already been extensively studied in the context of holographic entanglement measures, e.g., \cite{Lopez,Albash:2010mv,Balasubramanian:2010ce,Liu:2013iza,Casini:2015zua,Alishahiha:2014cwa,Fonda:2014ula,Leichenauer:2015xra,Alishahiha:2014jxa}. It was argued in \cite{Liu:2013iza} that the evolution of EE can be captured by the picture of an entanglement tsunami, \textit{i.e.}, a wave propagating inward from the boundary of the entangled region. In this setup, one can derive in detail several universal features in the evolution of nonlocal measures in quenched holographic systems. In particular, for large entangling regions, the evolution of EE experiences different regimes of an early time quadratic growth, an intermediate linear-growth, and a late time saturation. Further, employing the same picture, the nonequilibrium evolution of other measures such as mutual and tripartite information has been considered in \cite{Leichenauer:2015xra,Alishahiha:2014jxa}. Related studies attempting to better understand the evolution of entanglement measures in more general time-dependent bulk geometries have also appeared in \cite{Alishahiha:2014cwa,Fonda:2014ula}.

In \cite{BabaeiVelni:2020wfl}, we investigated the holographic proposals concerning the EWCS for Vaidya geometries describing a thin shell of null matter collapsing into the AdS$_{d+1}$ vacuum to form a one-sided black-brane. We considered a symmetric configuration consisting of two disjoint strips with equal width. A surprising result we found was that for large entangling regions, the evolution of the EWCS experiences the same scaling regimes as EE and the rate of growth of $\EW$ is equal to $S_A$. In $2+1$ dimensions, we presented a combination of numerical and analytic results which support this behavior. Moreover, in higher-dimensional cases we provided a numerical treatment and examine the various regimes in the growth of the EWCS. Despite numerical results for the time dependence of this quantity, the question of full time evolution and in particular the rate of growth at intermediate times has not been thoroughly investigated. Hence, in the present paper, we employ an analytic treatment to study the full time evolution of EWCS in higher-dimensional Vaidya geometries. While this analysis can be done for a generic asymptotically AdS spacetime, in order to gain better insight into the properties of $\EW$, we generalize our study to specific nonrelativistic boundary theories, in particular, those with Lifshitz and hyperscaling violating exponents. 

It is worthwhile to mention that as previously noted in \cite{Alishahiha:2014cwa,Fonda:2014ula}, a nontrivial dynamical exponent can affect the rate of entanglement growth and hence the corresponding saturation time. We will confirm that (up to a time shift) the qualitative behavior of HEE and boundary measures dual to the EWCS is similar even in these nonconformal theories. 
Nevertheless, there are some challenges associated with studying holographic duality in nonrelativistic backgrounds, see e.g., \cite{Keeler:2013msa,Gentle:2015cfp}. In particular, the validity of HRT prescription and the ability to construct a well-defined entanglement wedge in the Lifshitz background is questionable \cite{Gentle:2015cfp}. In the present paper, we naively assume that the standard prescriptions for computing holographic entanglement entropy and the definition of the entanglement wedge are valid even in Lifshitz geometry.

This paper is structured as follows: In Section \ref{sec:setup} we introduce the required
preliminaries. We briefly review the Vaidya background with Lifshitz and hyperscaling violating exponents and then we
carefully examine the general form of
the EWCS functional in this geometry. Then, in section \ref{sec:AdS} we study in detail the evolution of the EWCS in the formation
of a black-brane modeled by the time-dependent geometry for a null shell collapsing into the
AdS vacuum spacetime. Throughout this section, we consider the large subregion limit, where analytic formulas for different information measures can be found. Afterwards, in section \ref{sec:EWCSinLHS} we focus on the case where the boundary theory is nonrelativistic and we apply the same prescription to compute the evolution of the EWCS. We present an analytic
derivation of the effect of Lifshitz and hyperscaling-violating exponents on entanglement velocity and the early and late-time behavior of the EWCS. Finally, we discuss some implications of
our results, as well as possible future directions, in section \ref{sec:results}.
%%%%%%%%%%%%%%%%%%%%%%%%%%%%%%%
\section{Setup}\label{sec:setup}
%%%%%%%%%%%%%%%%%%%%%%%%%%%%%%%
We are interested in the time evolution of the EWCS in the presence of an infalling thin null shell in a background with Lifshitz and hyperscaling-violating exponents. In this section, we review the background metric and then obtain the EWCS profile for two disjoint strip regions in the connected phase (see figure \ref{fig:regions}). 
\subsection{Vaidya Geometry with Lifshitz and Hyperscaling Exponents} 
In this paper, we consider the following metric which describes collapsing of a null shell and the formation of a black-brane in the vacuum background with Lifshitz scaling $z$ and hyperscaling-violating exponent $\theta$
\begin{equation}\label{metric}
\begin{split}
	ds^2&=r^{-2\frac{D-\theta}{D}}\qty(-\frac{f(r,v)}{r^{2z-2}}\dd{v}^2-\frac{2}{r^{z-1}}\dd{v} \dd{r}+\dd{\vb{x}}_{D}^2),
\\
 f(r&,v)=\begin{cases}
	1& v<0\\
	g(r) &v>0
\end{cases}, \qquad   
g(r)=1-\left(\frac{r}{\rh}\right)^{D-\theta+z}, 
\end{split}
\end{equation}
where we employ the Eddington-Finkelstein like coordinate, \textit{i.e.}, $\dd{v}=\dd{t}-r^{z-1}\frac{\dd{r}}{f } $. 
 One may note that inside the shell ($v<0$) the spacetime is given by hyperscaling-violation Lifshitz (HSL)  metric while the outside region ($v>0$)  is an  asymptotically HSL black-brane. As we mentioned it may model a sort of global quench in the dual boundary theory. This metric reduces to the relativistic AdS-Vaidya geometry for $z=1$ and $\theta=0$.  
We assume that this Vaidya-type metric is a solution to Einstein's theory with some suitable matter fields. For example, it was shown that the Einstein-Maxwell-dilaton theory admits such solution \cite{Alishahiha:2012qu}
\begin{equation}
	\mathcal{I}=-\frac{1}{16\pi G_{N}}\int \dd[D+2]{x}\qty[R-\frac{1}{2}(\partial \phi)^2+V_0 e^{\gamma \phi}-\frac{1}{4}\sum_{i}^{2}e^{\lambda_{i}\phi}F_{i}^2].
\end{equation}
It is appropriate to define an effective dimension $$d:=D-\theta+1$$ that for $\theta=0$ reduces to the dimension of dual boundary QFT$_{d}$. 

At late times, the bulk metric \eqref{metric} describes a black-brane with a horizon at $\rh$. The temperature,  thermal entropy density  associated with the horizon, and energy density \cite{Alishahiha:2019lng} are
\begin{equation}\label{eq:TSE}
	T=\frac{d+z-1}{4\pi \rh^{z}}, \qquad \mathcal{S}_{\text{th}}=\frac{1}{4G_N}\frac{1}{\rh^{d-1}}, \qquad \mathcal{E}=\frac{d-1}{16\pi G_N}\frac{1}{\rh^{d+z-1}}.
\end{equation}
In what follows we assume that $z\geq1$ and $d>1$. It ensures that the null energy condition is satisfied and the black-brane solution is thermodynamically stable. Also in this case the zero-temperature entanglement entropy is consistent with the previous results \cite{Dong:2012se}. 
%%%%%%%%%%%%%%%%%%%%%%%%%%%%%%%
\subsection{EWCS in Vaidya Background}
\begin{figure}
	\begin{center}
			\includegraphics[scale=0.8]{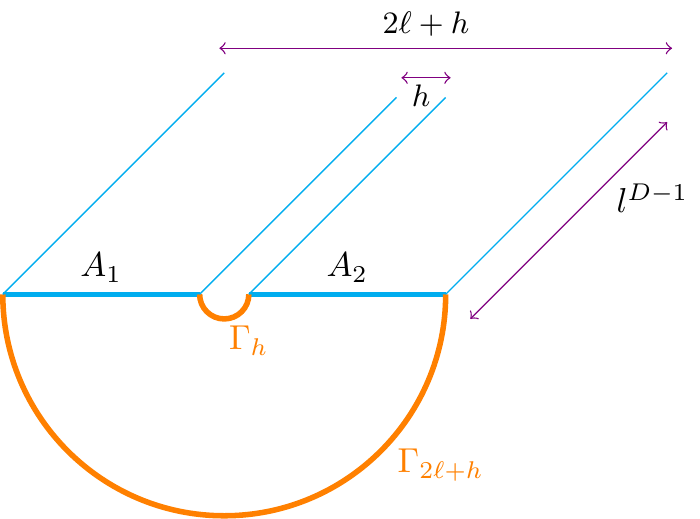}
		\hspace*{0.7cm}
		\includegraphics[scale=0.8]{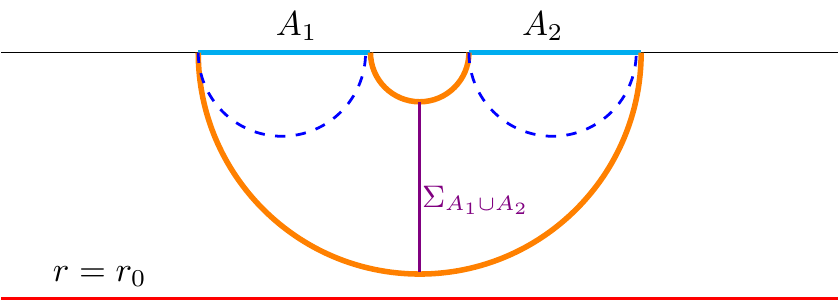}
	\end{center}
	\caption{Schematic configuration for computing HMI (left) and EWCS (right). Here we just demonstrate the connected configuration where both the HMI and EWCS are nonzero.}
	\label{fig:regions}
\end{figure}

%%%%%%%%%%%%%%%%%%%%%%%%%%%%%%%
We are interested in the time evolution of EWCS associated with two identical long strips ($A_1$ and $A_2$) on the boundary (see figure \ref{fig:regions}). We will take them to be of width $\ell$, length  $\tL$ and are separated by distant $h$.  The EWCS is given by eq. \eqref{eq:EwCS} and we assume that $h\ll \ell$, ensures the entanglement wedge is connected and the EWCS does not vanish. In this configuration, the entanglement wedge is bounded by two extremal hypersurfaces anchored to boundary strips with size $2\ell+h$ and $h$. In figure \ref{fig:regions} we denote these hypersurfaces by $\Gamma_{2\ell+h}$ and $\Gamma_{h}$, respectively. Moreover, we denote the turning points of $\Gamma_{2\ell+h}$ and $\Gamma_{h}$ by $r_u$ and $r_d$ respectively.
The EWCS can be parameterized as $v=v(r)$ and due to the reflection symmetry about $x=0$ the induced metric on $\Sigma_{A_1
\cup A_2}$ simplified as follows
\bea\label{inducedEWCS}
ds_{\rm ind.}^2=r^{\frac{-2(D-\theta)}{D}}\left[-\left(f(r,v){v'}^2+2r^{z-1}v'\right)\frac{\dd{r}^2}{r^{2z-2}} +\dd{\vb{x}}_{D-1}^2\right].
\eea
Now using eq. \eqref{eq:EwCS} the $\EW$ which is proportional to the extremal area of the corresponding codimension-2 hypersurface is  given by the following functional
\bea
\EW=\frac{\tL^{D-1}}{4G_N}\int \dd{r}\frac{\mathpzc{E}}{r^{d-1}},\qquad \mathpzc{E}:=\sqrt{-\frac{2v'}{r^{z-1}}-\frac{f(r,v){v'}^2}{r^{2z-2}}}.
\eea
Extremizing the above expression yields the equation of motion for $v(r)$
\bea\label{EOMEWCS}
\frac{\partial}{\partial r}\left(\frac{r^{z-1}+f(r,v) v'}{r^{d+2z-3}\mathpzc{E}}\right)=\frac{{v'}^2}{2r^{d+2z-3}\mathpzc{E}}\frac{\partial f}{\partial v}.
\eea
To obtain the extremal hypersurface we need to solve the above equation by assuming certain boundary conditions 
\bea\label{bdyconditionEWCS}
v(r_d)=v_d, \qquad v(r_u)=v_u,
\eea
where $r_d$ and $r_u$ are the turning points of hypersurfaces $\Gamma_{h}$ and $\Gamma_{h+2\ell}$ respectively. Note that for $v\neq 0$,  $\frac{\partial f}{\partial v}=0$ and we have a conserved quantity as follows
\begin{equation}
    \frac{r^{z-1}+f(r,v) v'}{r^{d+2z-3}\mathpzc{E}}=\frac{\Q}{\rh^{d+z-2}}.
\end{equation}
Using this fact, we can solve eq. \eqref{EOMEWCS} for $v'(r)$ to get
%    \begin{equation*}
% 	{v'}_{\pm}=-\frac{r^{z-1}}{f}\qty(1\mp\frac{(\frac{\Q}{\rh^{d+z-2}})\;r^{d+z-2}}{\A}).
% \end{equation*}
% \begin{equation}\label{vpm}
% 	{v'}_{\pm}=-\frac{r^{z-1}}{f}\qty(1\mp\frac{\Q\;\qty(\frac{r}{\rh})^{d+z-2}}{\A}), \qquad \A:=\sqrt{f+\Q^2(r/\rh)^{2(d+z-2)}}.
% \end{equation}
\begin{equation}\label{vpm}
	{v'}_{\pm}=-\frac{r^{z-1}}{f}\qty(1\mp\frac{\Q\;\qty(r/\rh)^{d+z-2}}{\A}), \qquad \A:=\sqrt{f+\Q^2(r/\rh)^{2(d+z-2)}}.
\end{equation}
One may note that $\Q \to -\Q$ implies $v_{\pm} \to v_{\mp}$, so  in what follows without loss of generality, we only consider the $v_{-}$ branch. 
Discontinuity of $f(r,v)$ requires studying $\Q$ in the  HSL region ($v<0$) and black-brane region ($v>0$) separately and then matching the results at the point where the null shell and the EWCS intersect.  Note that in the following we will use  the subscripts $a$ and $b$ to refer to quantities on the HSL and black-brane side, respectively.
\begin{figure}
	\begin{center}
		\includegraphics[scale=1]{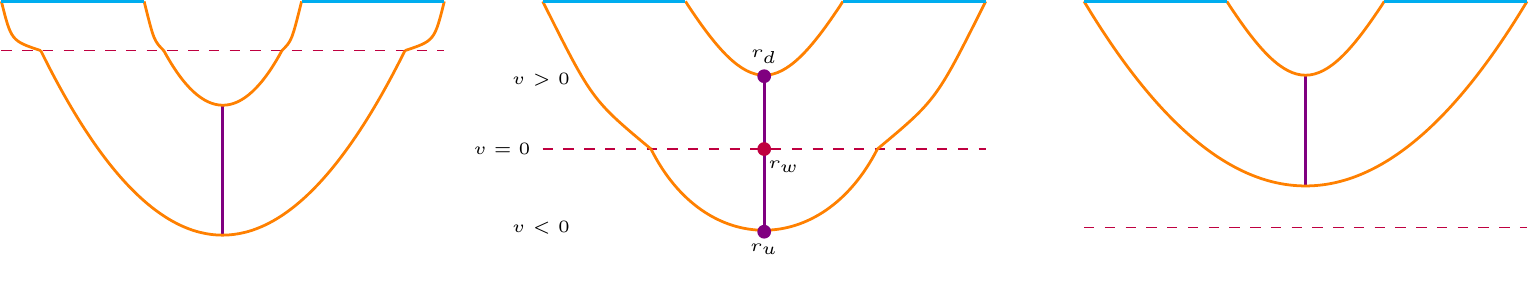}
	\end{center}
\caption{Schematic configurations for HRT and EWCS hypersurfaces corresponding to the connected phase where the separation between the boundary entangling regions is small. Outside the collapsing shell (indicated in dashed purple), i.e, $v>0$, the hypersurfaces propagate in an AdS black-brane spacetime. On the other hand, in $v<0$ region, they propagate in a pure AdS geometry. \textit{Left}: At early times the null shell lies near the boundary of the spacetime and $\Sigma$ does not intersect the shell. \textit{Middle}: During intermediate stages of evolution $\Sigma$ crosses the null shell. \textit{Right}: At the late-time $\Sigma$ lies entirely in black-brane geometry.}
	\label{fig:3cases}
\end{figure}
\subsubsection*{HSL region ($v<0$)}
In this case $f(r,v)=1$ and eq. \eqref{vpm} reduces to 
\begin{equation} 	
	v_{\;\Minus}'=-r^{z-1}\left(1+\frac{\Qa r^{d+z-2}}{\Aa}\right),
\end{equation}
where $\Qa \rh^{d+z-2}:=\Q$ and $\Aa:=\A(\Qa \rh^{d+z-2},r)\eval_{f=1}$.
Therefore the profile of EWCS becomes
\begin{align}
	v_{\; \Minus}(r)=c_{\; \Minus}-\frac{r^z}{z}- \F(r),
\end{align}
where
% \begin{equation}\label{eq:F}
% 	\F(r):=\int 	\frac{\Qa r^{d+2z-3}\; \dd{r}}{\sqrt{1+\Qa^2 r^{2(d+z-2)}}}=	\frac{r^z}{z} \; {_2F_1}\left(\frac{1}{2},\frac{z}{2(2-d-z)},1+\frac{z}{2(2-d-z)},\frac{-r^{2(2-d-z)}}{\Qa^2}\right).
% \end{equation}
\begin{equation}\label{eq:F}
\F(r):=\int^r 	\frac{\Qa {r'}^{d+2z-3}\; \dd{r'}}{\Aa}=	\frac{r^z}{z} \; {_2F_1}\qty(\frac{1}{2},\mathpzc{b},1+\mathpzc{b},-\frac{r^{z/\mathpzc{b}}}{\Qa^2}), 
\end{equation}
and $\mathpzc{b}:=\frac{z}{2(2-d-z)}$.
By imposing the boundary condition $v_{\; \Minus}(r_u)=v_u$  we can fix the integration constant as 
\bea
c_{\; \Minus}=v_u+\frac{r_u^z}{z}+ \F(r_u).
\eea
Note that for $d=2$ and $z=1$ the above expressions precisely match with the previous results reported in \cite{BabaeiVelni:2020wfl}.

\subsubsection*{black-brane region ($v>0$)}
Now, let us consider the profile in the black-brane region where $f(r,v)=g(r)$. In this case using eq. \eqref{vpm}, one derives the profile of EWCS in $v>0$ region as follows
\bea
v_{\;\Plus}(r)=c_{\;\Plus}-\G(r)
-\Qb \rh^{z}\int_{\frac{r_d}{\rh}}^{\frac{r}{\rh}}\dd{u}\frac{u^{d+2z-3}}{g(u)\Ab}, \quad u:=\frac{r}{\rh},
\eea
where $\Ab=\A(\Qb,r)\eval_{f=g}$  and  $\mathcal{G
}$ is defined as 
% \begin{equation}
% 	\G(r):=\int \frac{\dd{r}}{g(r)}=\frac{r^z}{z}\; {_2F_1}\left(1, \frac{z}{d+z-1}, 1+\frac{z}{d+z-1},\left(\frac{r}{\rh}\right)^{d+z-1}\right).
% \end{equation}
\begin{equation}
	\G(r):=\int \frac{\dd{r}}{g(r)}=r \; {_2F_1}\qty(1,\bar{\mathpzc{b}}, 1+\bar{\mathpzc{b}},\qty(\frac{r}{\rh})^{1/\bar{\mathpzc{b}}}),
\end{equation}
where $\bar{\mathpzc{b}}:=\frac{1}{d+z-1}$.
Once again using the boundary condition $v_{\; \Plus}(r_d)=v_d$ one can obtain the integration constant
\bea
c_{\; \Plus}=v_d+\G(r_d).
\eea

\subsubsection*{Matching at $v=0$}
By integrating eq. \eqref{EOMEWCS} across the null shell at $v(r_w)=0$, we obtain the corresponding matching condition
\bea\label{matchd}
\dv{r}{v}\eval_{\; \Minus}-\dv{r}{v}\eval_{\; \Plus}=-\frac{1}{2}\left(\frac{r_w^{d}}{\rh^{d+z-1}}\right).
\eea
Solving the above condition, we get the relation between the conserved quantities, \textit{i.e.}, $\Qa$ and $\Qb$ as follows 
\begin{align}\label{eq:Qb}
		\Qb&=\pm\frac{\left(r_w^{d+z-1}-2\rh^{d+z-1}\right)\Qa+r_w \Aa(r_w)}{2\rh}.
\end{align}
Without loss of generality in the rest of the paper, we take the positive sign. On the other hand, the continuity of the EWCS profile at the intersection point, \textit{i.e.}, $v_a(r_w)=0=v_b(r_w)$ implies
\begin{subequations}
\begin{gather}
	v_u+\frac{r_u^{z}}{z}+\F(r_u)-\frac{r_w^{z}}{z}-\F(r_w)=0 ,
 \label{eq:match1}\\
	v_d+\G(r_d)-\G(r_w)
	-\Qb \rh^{z}\;\int_{\frac{r_d}{\rh}}^{\frac{r}{\rh}}\dd{u}\frac{ u^{d+2z-3}}{g(u)\Ab}=0. \label{match2}
\end{gather}
\end{subequations}
Note that when $r_d $ is located in the black-brane region where $v_d>0$ one can use $\dd{t}=\dd{v}+r^{z-1}\frac{\dd{r}}{g}$ and $t=v_d+\G(r_d)$ to express the boundary time in terms of other parameters as follows
\begin{equation}\label{eq:tboundary}	
t=\G(r_w)+\Qb \rh^{z}\;\int_{\frac{r_d}{\rh}}^{\frac{r_w}{\rh}}\dd{u}\;\frac{ u^{d+2z-3}}{g(u)\Ab}.
\end{equation}
These conditions are sufficient to obtain EWCS in terms of $v_u$, $v_d$ as well as $r_w$ which are related to the boundary parameters $\ell$, $h$, and $t$, respectively. So in principle, by using them one may obtain the contribution to the EWCS in both regions
\begin{subequations}
\begin{gather}
	\tEWa=
\int_{r_w}^{r_u}\frac{r^{1-d} \; \dd{r}}{\Aa} ,
 \label{eq:EWLif}\\
	%=\frac{\tL^{D-1}}{4(2-d)G_N} \qty(r^{2-d}{_2F_1}\left(\frac{1}{2},\frac{2-d}{2(d+z-2)},1+\frac{2-d}{2(d+z-2)},-\Qa^2 r^{2(d+z-2)}\right)) \eval_{r_w}^{r_u}, \\
	\tEWb=\frac{1}{\rh^{d-2}}\int_{\frac{r_d}{\rh}}^{\frac{r_w}{\rh}} \frac{u^{1-d}\; \dd{u}}{\Ab},
 %=\frac{\tL^{D-1}}{4G_N}\frac{1}{\rh^{d-2}} \int_{\frac{r_d}{\rh}}^{\frac{r_w}{\rh}} \frac{u^{1-d}\; \dd{u}}{\sqrt{g(u)+\Qb^2u ^{2(d+z-2)}}}
	\label{eq:EWBH}
\end{gather}
\end{subequations}
where we have defined $\EW:=\frac{\tL^{D-1}}{4G_{N}}\tEW$ for simplicity. Finally using these equations we can read $\EW$ as 
\bea
\EW=\EWa+\EWb.
\eea
In the next sections following \cite{Liu:2013qca}, we will study the time evolution of $\EW$ in three different scaling regimes. Before we proceed further, let us comment on an assumption that greatly simplifies the calculation.

\subsection*{Notes on $\Qa$ and $\Qb$}
\begin{figure}
	\begin{center}
		\includegraphics[scale=0.8]{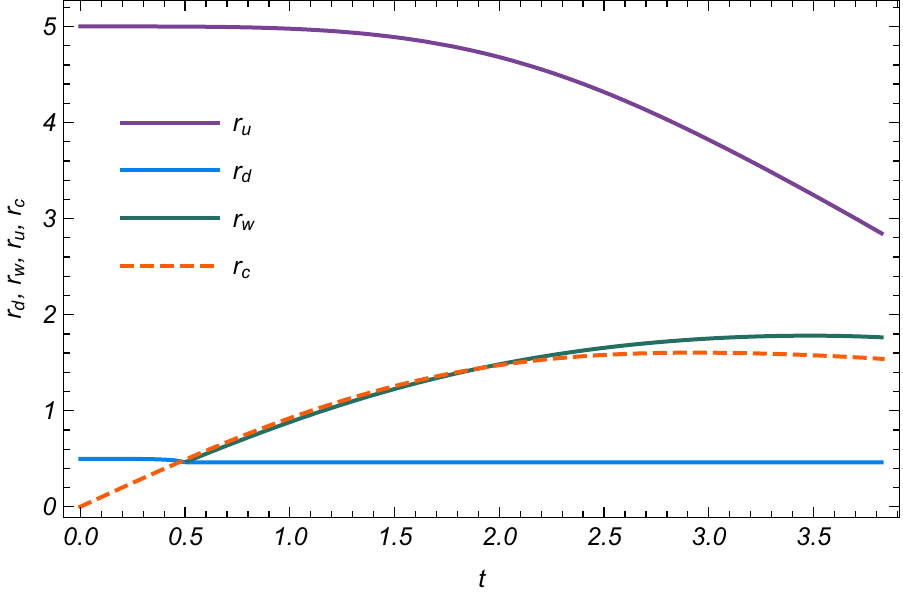}
		\hspace*{.2cm}
		\includegraphics[scale=0.8]{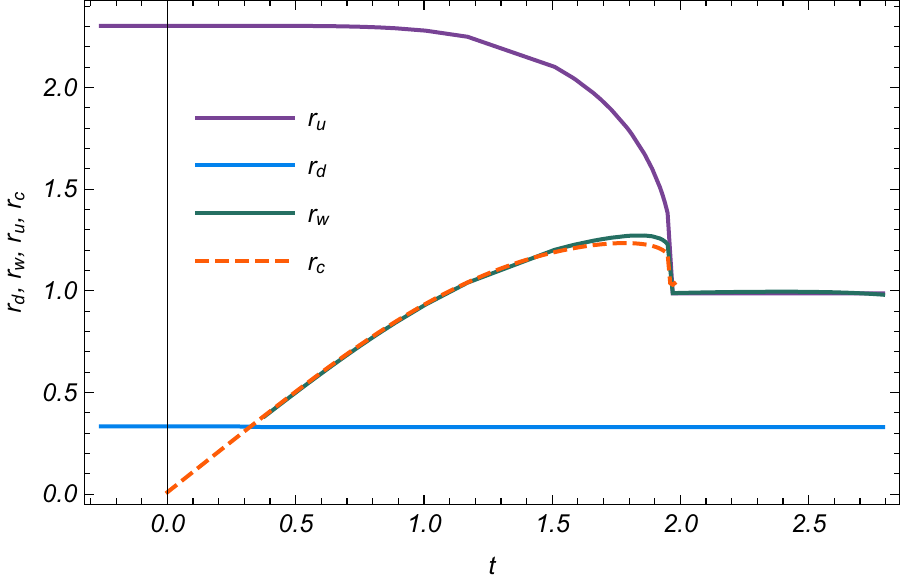}
	\end{center}
	\caption{
Time evolution of $r_d$, $r_u$, $r_w$ and $r_c$ for $z=1$. Clearly, at the early time we have $r_w\approx r_c$ and the deviation between them becomes more pronounced near the saturation time. \emph{Left:} $d=2, h=1$, and $\ell=4.5$. \emph{Right:} $d=3, h=0.4$, and $\ell =1.18$. Here we set $\rh=1$.}
	\label{fig:rdrwrurcnumz1}
\end{figure}

In what follows we mainly consider a specific limit where the characteristic size of the boundary entangling regions is large compared to the inverse temperature, \textit{i.e.}, $\ell T^{1/z}\gg h T^{1/z}\gg 1$. Indeed, we will argue that $\Qa \approx 0$ in this limit. It leads to a great simplification in the semi-analytic results for $\EW$. To show this, first, note the relation between $r_w$ and $r_c$. These quantities indicate positions where the null shell intersects $\Sigma_{A_1\cup A_2}$ and $\Gamma_{2\ell+h}$ respectively (see figure \ref{fig:3cases}). It is easy to show that $r_c^{z}=z v_u+r_u^{z}$ which simplifies eq. \eqref{eq:match1} as follows 
\begin{equation*}
    r_w^{z}-r_c^{z}=	z\F(r_u)-z\F(r_w).
\end{equation*}
Note that the null shell intersects EWCS at $r_w$ and in general $r_w\neq r_c$  but one may expect that $r_c \approx r_w$. Indeed, our numerical results confirm this expectation see, \textit{e.g.,} figure \ref{fig:rdrwrurcnumz1}. Based on this figure, when the connected configuration
is always favored for any boundary time, these quantities precisely match at early and intermediate times. On the other hand, when the disconnected configuration is favored at late times, although at early times $r_c$ matches with $r_w$, the deviation between them becomes more pronounced as time evolves. Further, by expanding the hypergeometric function in eq. \eqref{eq:F}, it is easy to show that
	\begin{align}
	\frac{r_w^z}{z}-\frac{r_c^z}{z}&=\sum_{n=0}^{\infty}\mathcal{C}_{\; \;n}^{-\frac{1}{2}}\frac{r_u^{p_n}-r_w^{p_n}}{p_n}\Qa^{2n+1}, \quad p_n=2n(d-1)+d+2(z-1)(n+1),
	%&=\qty(\frac{r_u^{d+2z-2}-r_w^{d+2z-2}}{{d+2z-2}})\Qa-\qty(\frac{r_u^{3(d-2)+4z}-r_w^{3(d-2)+4z}}{2({3(d-2)+4z})})\Qa^3+\order{\Qa^5}.
\end{align}
where  $\mathcal{C}_{\; \;n}^{-\frac{1}{2}}=\Gamma(1/2)/\Gamma(n+1)\Gamma(1/2-n)$. So for $r_c \approx r_w\ll r_u$  this equation requires $\Qa \to 0$.
Then from eq. \eqref{eq:Qb} we get
\begin{equation}\label{eq:QaQb}
	\Qa \approx 0 \qand 	\Qb \approx -\frac{r_w}{2\rh}, \qquad (r_w \ll r_u).
\end{equation}
In the following sections, we will employ these relations to study the scaling of $\EW$ during the early and intermediate stages of time evolution.

However when $r_w \approx r_u$, we cannot assume $\Qa=0$. Indeed, in this situation, $\Qb\approx0$ and $\Qa \neq 0$. To see this, note that as $r_w\to r_u$ the $\Gamma_{2\ell+h}$ settles down to its final static configuration as it lies in the black-brane geometry and hence $t\approx v_{u}+\G(r_u)$. On the other hand, the latter should be consistent with eq. \eqref{eq:tboundary} upon
substituting $r_w\approx r_u$ 
\begin{equation}
v_u+\G(r_u)=\G(r_u)+\Qb \rh^{z} \int_{\frac{r_d}{\rh}}^{\frac{r_u}{\rh}}\dd{u}\;\frac{  u^{d+2z-3}}{g(u)\Ab}.
\end{equation}
Now, we note that $v_u=v(r_u \to r_w)\approx 0$ and then the above equation implies $\Qb \to 0$. In this situation, by solving eq. \eqref{eq:Qb} for $\Qb=0$ one obtains
\begin{equation}\label{eq:QaQblatetime}
\Qa^2\approx  \frac{r_u^{2}}{4 \rh^{d+z-1}\qty(\rh^{d+z-1}-r_u^{d+z-1})}, \qand \Qb\approx 0, \qquad (r_w \approx r_u).
\end{equation}
As we will argue, this result allows us to explore the saturation of $\EW$ as the EWCS lies entirely in the black-brane region.

\section{EWCS in AdS black-brane ($z=1,\theta=0,d>2$)} \label{sec:AdS}
We shall now discuss the time evolution of EWCS in the various regimes with a collapsing shell of matter in asymptotic AdS spacetime which describes a relativistic quantum quench. As we explained in the previous section, we are interested in the connected configuration where EWCS is nontrivial. We also focus on the regime where the null shell intersects the EWCS (see figure \ref{fig:3cases}).
In such circumstances, we investigate three different regimes: i) \emph{Early growth} describes the behavior of $\EW$ immediately after the null shell passes through the $r_d$. So in this regime, we should assume $r_d\lesssim r_w$.  ii) \emph{linear-growth} characterizes situation where $r_u\gg r_w \gg r_d$. iii) \emph{Saturation} which shows how the $\EW$ reaches to its equilibrium value.

 We remind the reader that a similar study for $d=2$ has been done in \cite{BabaeiVelni:2020wfl}, thus here we just focus on $d\ge3$. Moreover, in the next section, we discuss the effects of $z$ and $\theta$ on the EWCS evolution by exploring the HSL background.

\subsection{Early growth}
As we just mentioned, 
the collapsing shell does not intersect $\Sigma$ at very early time (see figure \ref{fig:3cases}).  Although, the entanglement entropy starts its early growth during this period of time \cite{Liu:2013qca}, the
 EWCS  does not change and is given by $\EWa$. So here, by  early growth we mean the time evolution of EWCS right after the null shell intersects $\Sigma$.
 To compute the time dependence of $\EW$ at early time, we suppose $r_d \lesssim r_w$ and $r_d \ll \rh \ll r_u $. It allows us to consider $r_w=r_d +\delta$ where $\delta \ll r_d \ll \rh$. Moreover, the large interval $\rh\ll r_u$ lets to use eq. \eqref{eq:QaQb} in order to expand eq. \eqref{eq:EWBH} up to the first order that depends on $\rh$
 \begin{align}\label{eq:earlybh}
	\tEWb=\frac{\delta}{r_d^{d-1}}-\frac{(d-1)}{2}\frac{\delta^2}{r_d^{d}}
 +\frac{\delta^2+2r_d \delta}{4 \rh^{d}}+	\order{\frac{\delta^3}{r_d^{d+1}},\frac{\delta\; r_d^{d+1}}{\rh^{2d}}}.
\end{align}
 On the other hand, we may obtain $\EWa$ piece by applying $\Qa=0$ in eq. \eqref{eq:EWLif} 
\begin{align}\label{eq:earlyads}
	\tEWa&=\frac{1}{(d-2)r_w^{d-2}}-\frac{1}{(d-2)r_u^{d-2}}+\order{\Qa^{\frac{d-2}{d-1}}}.
\end{align}
Further, eq. \eqref{eq:earlybh} suggests that one should expand $\EWa$ around $r_w=r_d+\delta$ up to $\order{\delta^{3}}$ 
\begin{equation} \label{eq:earlyads_rd}
	\tEWa	=\frac{1}{(d-2)r_d^{d-2}}-\frac{1}{(d-2)r_u^{d-2}}-\frac{\delta}{r_{d}^{d-1}}+\frac{(d-1)}{2}\frac{\delta^2}{r_d^d}+\order{\delta^3}+\order{\Qa^{\frac{d-2}{d-1}}}.
\end{equation}
Now we are equipped with all we need to calculate the early growth of EWCS in terms of $\delta$ as follows 
\begin{equation}\label{eq:EWdelta} 
	\tEW=\frac{1}{(d-2)r_d^{d-2}}-\frac{1}{(d-2)r_u^{d-2}}+
	\frac{\delta^2+2 r_d \delta}{4\rh^d}+	\order{\delta^3,r_d^{d}} .
\end{equation}
However,  we are interested in the explicit dependence of the EWCS on boundary time. Therefore we expand eq. \eqref{eq:tboundary} to get  $\delta$ in the terms of $t$ 
\begin{align}\label{delta_t}
t=\delta+r_d+\order{r_d^{d},r_d^{d-1} \delta^2}.
\end{align}
Substituting the above result into eq. \eqref{eq:EWdelta} we obtain the time scaling of $\EW$ in this regime as
\begin{equation}\label{linearz1}
	\tEW \approx\tEWa(r_d,r_u)+\frac{t^{2}-r_d^2}{4\rh^{d}}.
\end{equation}
The first term describes the $\tEW$ in the static AdS space and so one may call it  the  vacuum contribution. By subtracting it we obtain one of our main results
\begin{equation}\label{eq:early-AdS}
	\eqbox{\Delta	\EW \approx \frac{\pi\tL^{d-2} \mathcal{E}}{d-1}  \qty(t^{2}-\frac{h^2}{\mathpzc{c}^2}) ,}
\end{equation}
where the energy density of the boundary state $\mathcal{E}$ is defined in eq. \eqref{eq:TSE}. We also use $h\approx \mathpzc{c} r_d$ to fully express the final result in terms of the boundary quantities \footnote{
For  $r_d\ll\rh$ (low temperature limit $h T^{1/z}\ll 1$)
\begin{equation}\label{eq:h-rd}
r_d\approx\frac{h}{\mathpzc{c}}\qty(1-\frac{\sqrt{\pi}}{(d+z)\mathpzc{c}^{d+z}}\frac{\Gamma\left(\frac{2d+z-1}{2 (d-1)}\right)}{\Gamma\left(\frac{d+z}{2(d-1)}\right)}\left(\frac{h}{r_0}\right)^{d+z-1})
, \quad \mathpzc{c}=\frac{2\sqrt{\pi}\,\Gamma \left(\frac{d}{2(d-1)}\right)}{\Gamma \left(\frac{1}{2(d-1)}\right)}.
\end{equation}
}. 
The above result shows that the early growth  of $\EW$ starts at $t\sim h$, in contrast to entanglement entropy where its early growth begins at $t=0$.  Moreover, as one may expect, for $h\to 0$ this result matches with the early growth of entanglement entropy \cite{Liu:2013qca}.

\subsection{Linear-growth and saturation}\label{subsec:Lieanr}

\begin{figure}
	\begin{center}
		\includegraphics[scale=.9]{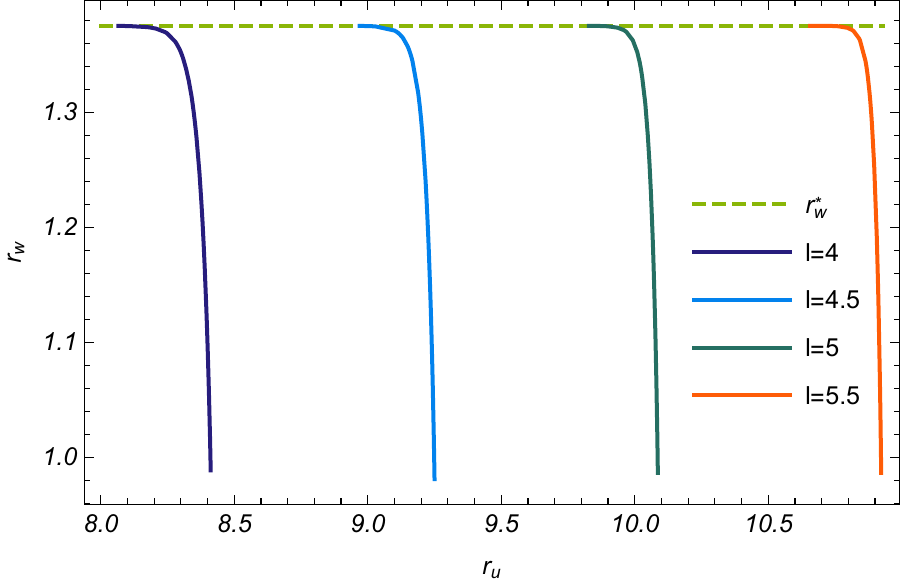}
			\end{center}
	\caption{$\left(r_w(t), r_u(t)\right)$ for different values of $\ell$ with $h=2.1$ and $z=1$. During the evolution, $r_w$ approaches $r_w^{\star}$ corresponds to the linear-growth regime. For these values of $\ell$ and $h$ the disconnected configuration is favored at late times and the extremal hypersurface jumps at some saturation time where each curve stops. Here we set $\rh=1$.}
	\label{fig:rwstarz1}
\end{figure}

Here we survey regime corresponds to $h\ll \rh\ll t\ll \ell$ where  $r_w\ll r_u$. First, let us recall eqs. \eqref{eq:EWBH} and \eqref{eq:tboundary} as 
\begin{equation}\label{EWt}
	\begin{gathered}
		\tEWb=\frac{1}{\rh^{d-2}} \int_{\frac{r_d}{\rh}}^{\frac{r_w}{\rh}} \frac{u^{1-d}\; \dd{u}}{\Ab(u,u_w)},\qquad
			t=\G(r_w)+\Qb \rh  \int_{\frac{r_d}{\rh}}^{\frac{r_w}{\rh}} \frac{u^{d-1} \dd{u}}{g(u)\Ab(u,u_w)},
   \end{gathered}
\end{equation}
where $\Ab$ depends on $u_w$ via eq. \eqref{eq:Qb}.
 Now let us suppose that there is a minimum for $\Ab$ at $\um$ and so $\partial_{u}\Ab(u)\eval_{\um}=0$. Moreover, we also assume that at $u=\um$ there is a $u_w=\uws$ such that $\Ab(\um,\uws)=0$. Indeed, our numerical results confirm these assumptions see \textit{e.g.,} figure \ref{fig:rwstarz1}. Based on this figure, we see that during the evolution, $r_w$ approaches $r_w^{\star}$ corresponds to the linear-growth regime. In this plot, we consider the values of $\ell$ and $h$ such that the disconnected configuration is favored at late times. Hence the saturation is discontinuous and the corresponding extremal hypersurface jumps at some saturation time where each curve stops. Moreover, it is clear that, as $\ell$ increases, also the linear regime increases. Based on these observations the dominant part of the integrals in eq. \eqref{EWt} comes from the region near $\um$ and $\uws$. It is similar to what happens in the time evolution of HEE as discussed in \cite{Liu:2013qca}.
 Existence of $\uws$ implies $\Qb^{\star}=\Qb\eval_{\uws}$. Thus, by expanding $\Ab$ about $u=\um$ and $\uws$ (or $\Qb$), we get
\begin{gather}\label{A2}
	\Ab\approx\CA_1\; (u-\um)^2+\CA_2\; (\Qb^{\star}-\Qb),
\end{gather} 
where
\begin{gather*}	\CA_1:=\frac{1}{2}\partial^{2}_{u}\Ab\eval_{\um,\uws}, \qquad
	\CA_2:=-\partial_{\Qb}\Ab\eval_{\um,\uws},\\
	\Qb^{\star}:= \um \frac{\sqrt{(d-2)d}}{2(d-1)}, \qquad  \um:=\qty(\frac{2(d-1)}{d-2})^{1/d}.
\end{gather*}
Using eq. \eqref{A2} we estimate the integral of $\EW$ as
\begin{equation}
    \tEWb=\frac{\um^{1-d}}{\rh^{d-2}}  \int_{\frac{r_d}{\rh}}^{\frac{r_w}{\rh}} \frac{\dd{u}}{\Ab}.
\end{equation}
On the other hand, inserting eq. \eqref{A2} in eq. \eqref{EWt} and
simplify the resultant equation yields  
\begin{align}
	t-t_{\star}&\approx  \frac{\Qb^{\star} \rh \um^{d-1}}{g(\um)}\int_{\frac{r_d}{\rh}}^{\frac{r_w}{\rh}} \frac{\dd{u}}{\Ab},%\frac{\dd{u}}{{\sqrt{\frac{1}{2}\A_2(u-\um)^2+b\epsilon}}},
\end{align}
where $t_{\star}:=\G(r_w^{\star})$.
The above results imply the linear-growth of $\EW$ in this regime
\begin{equation}\label{eq:linear-AdS}
   \eqbox{\Delta \EW\approx \tL^{d-2}  \mathcal{S}_{\text{th}}  V_{W}\; (t-t_{\star}). }
\end{equation}
where $\Delta \EW=\EW-\EWa(r_w^{\star},r_u)$
and then the velocity $V_{W}$ is given by
\begin{equation}
V_{W}:=\frac{g(\um)}{\Qb^{\star}\um^{2d-2}}= \qty(\frac{d-2}{2(d-1)})^{\frac{d-1}{d}}\sqrt{\frac{d}{d-2}}.
\end{equation}
One may note that the linear behavior is similar to the growth of entanglement entropy. In addition, the expression for $V_W$ is the same as the entanglement velocity defined in \cite{Liu:2013qca}.

As the null shell approaches the turning point $r_u$, EWCS falls into the balk brane region. So the value of $\EW$ should reduce to that in the static black-brane geometry. In other words, EWCS saturates its equilibrium value which is the same as $\EW$ at the thermal state. To see this, note that eq. \eqref{eq:EWLif} shows $\EWa \to 0$ as $r_w\to r_u$. In addition, we recall eq. \eqref{eq:QaQblatetime} in this regime and by employing $\Qb=0$ in eq. \eqref{eq:EWBH} we get the $\EW$ for the black-brane geometry \cite{BabaeiVelni:2019pkw}
\begin{equation}\label{equilib}
    \EW=\frac{\tL^{D-1}}{4G_{N}}\int_{r_d}^{r_u}\frac{r^{1-d}\dd{r}}{\sqrt{g(r)}}.
\end{equation}
It is worthwhile mentioning that the above analysis works when the connected configuration is always favored for any boundary time during the thermalization and the resultant $\EW$ continuously saturates to the final equilibrium value given by eq. \eqref{equilib}. However, we should
also remark that when the disconnected configuration is favored, the late-time behavior of the EWCS changes such that it displays a discontinuous transition and immediately saturates to zero.

\section{EWCS in LHS black-brane  ($z\neq 1,\theta \neq 0$)}\label{sec:EWCSinLHS}
In this section, we generalize our studies to holographic theories with general dynamical critical exponent $z$ and hyperscaling violation exponent
$\theta$. Employing the analysis we have outlined in the previous section we will be able to explore a variety of scaling regimes in the time evolution of EWCS. As we will see its behavior is qualitatively similar to HEE previously discussed in \cite{Alishahiha:2014cwa,Fonda:2014ula}.

\subsection{Early growth}

\begin{figure}
	\begin{center}
		\includegraphics[scale=0.8]{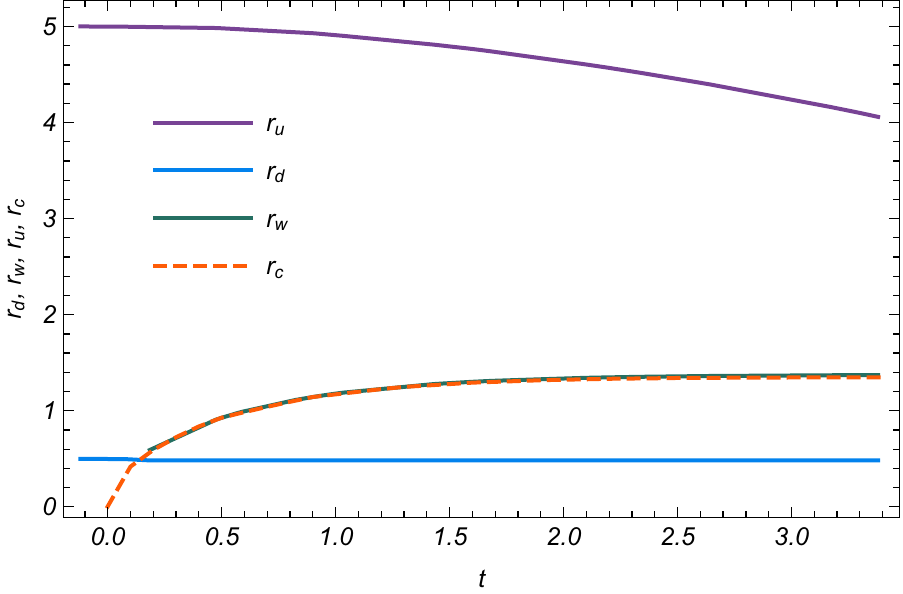}
		\hspace*{.1cm}
		\includegraphics[scale=0.8]{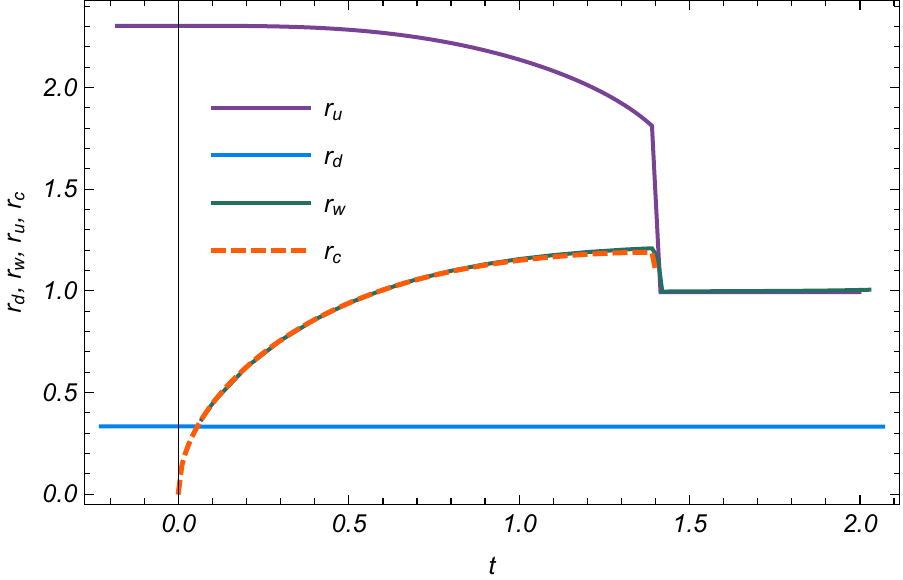}
	\end{center}
	\caption{
Evolution of $r_d$, $r_u$, $r_w$ and $r_c$ for $z=2$. Clearly, $r_w$ always coincides with $r_c$ for any boundary time. \emph{Left:} $d=2, h=1$, and $\ell=4.5$. \emph{Right:} $d=3, h=0.4$, and $\ell =1.18$. Here we set $\rh=1$.}
	\label{fig:rdrwrurcnumz2}
\end{figure}

As we explain in the previous section for $z=1$, at early times we expect that $r_w\approx r_c$. Indeed, our numerical results show that this behavior holds even in the non-relativistic case, see \textit{e.g.}, figure  \ref{fig:rdrwrurcnumz2}. 
In this case, $r_w$ always coincides with $r_c$ for any boundary time. Interestingly, comparing these results with the results depicted in figure \ref{fig:rdrwrurcnumz1},  we see that $r_w$ and $r_c$ will be in complete agreement if we choose larger values of $z$.
Recall that at early times, \textit{i.e.}, $t\ll \rh^z$, the shell does not reach $\Sigma$ which lies entirely in LHS geometry, and thus the
EWCS is a fixed constant given by the vacuum value. Let us start from eq. \eqref{eq:EWLif} and expand it up to the first term that depends on $\rh$
 \begin{align}
 	\tEWb& \approx \frac{1}{(d-2)r_{d}^{d-2}}-\sum_{n=0}^{z+1}\frac{\mathcal{C}_{n}^{2-d}}{d-2}\frac{\delta^{n}}{r_d^{d+n-2}}+\frac{\qty(r_d+\delta)^{z+1}-r_d^{z+1}}{2(z+1)\rh^{d+z-1}}.
\end{align}
where $\mathcal{C}_n^{m}$ denotes the binomial coefficient.
It suggests that one should expand eq. eq. \eqref{eq:EWBH} around $r_w=r_d+\delta$ up to $\order{\delta^{z+2}}$
\begin{equation}
\tEWa \approx\sum_{n=0}^{z+1}\frac{C_{n}^{2-d}}{d-2}\frac{\delta^{n}}{r_{d}^{d+n-2}}-\frac{1}{(d-2)r_u^{d-2}}%+\order{\frac{\delta^{z+2}}{r_d^{d+z}}}.
\end{equation}
Now using $\tEW=\tEWa+\tEWb$ we get 
\begin{equation}
    \tEW\approx \tEWa(r_d,r_u)+ \frac{\qty(r_d+\delta)^{z+1}-r_d^{z+1}}{2(z+1)\rh^{d+z-1}}.
\end{equation}
To obtain $\delta$ as a function of time we employ eq. \eqref{eq:tboundary} for $\{r_d,r_w\}\ll \{r_u, \rh\}$ and assume $r_w=r_d+\delta$ to read
\begin{equation}
\delta\approx (z t)^\frac{1}{z}-r_d.
\end{equation}
It allows us to express the time evolution as 
\begin{equation}\label{eq:early-Lif}
	 \Delta	\EW \approx\frac{2\pi\tL^{D-1} \mathcal{E}}{(d-1)(z+1)}  \qty((z t)^{1+\frac{1}{z}}-\qty(\frac{h}{\mathpzc{c}})^{z+1}),
\end{equation}
where $\mathpzc{c}$ is defined in eq. \eqref{eq:h-rd}. For $z=1$, this result reduces to that for Vaidya-AdS eq. \eqref{eq:early-AdS}. Moreover, one may note that the scaling at the early time just depends on the Lifshitz exponent $z$ and is independent of the hyperscaling-violating exponent $\theta$. It is worth noting that in $h \to 0$  limit, it reduces to HEE \cite{Alishahiha:2014cwa,Fonda:2014ula}.

 \subsection{linear-growth and saturation}

\begin{figure}
	\begin{center}
				\includegraphics[scale=.9]{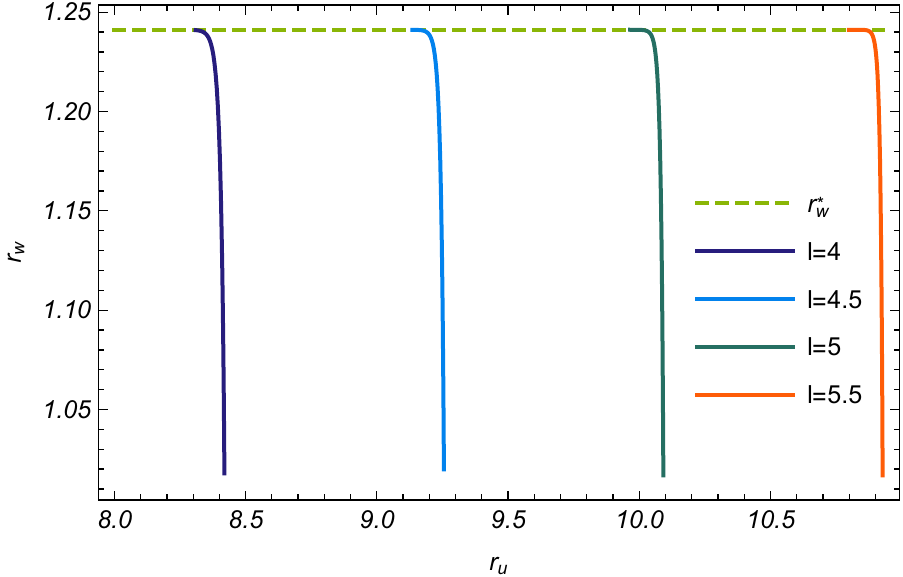}
	\end{center}
	\caption{$\left(r_w(t), r_u(t)\right)$ for different values of $\ell$ with $h=2.1$ and $z=2$. Similar to the relativistic case $r_w$ coincides with $r_w^{\star}$ in the linear-growth regime. The saturation at late times  is discontinuous. Here we set $\rh=1$.}
	\label{fig:rwstarz2}
\end{figure}
Here we follow the same steps as in section  \ref{subsec:Lieanr} to figure out time scaling of EWCS for $h\ll \rh\ll t^{1/z}\ll \ell$ where  $ \{r_w, r_c\}\ll r_u$. Again, in the presence of $\theta$ and $z$ we consider $\Ab$ that has a minimum at $\um$ and $\A(\um,\uws)=0$. Our numerical results
confirm these assumptions see e.g., figure \ref{fig:rwstarz2}. We see that during the evolution, $r_w$ approaches $r_w^{\star}$ in the linear-growth regime. Using the same expansion as eq. \eqref{A2} one finds
\begin{equation}
 \begin{split}
  \um=\qty(\frac{(d+z-2)}{d+z-3})^{\frac{1}{d+z-1}},\quad
		\Qb^\star= - \um\frac{\sqrt{(d+z-3)(d+z-1)}}{2(d+z-2)}.
\end{split}
\end{equation}
Once again, these assumptions simplify  expressions for $\EWb$ and $t-t_{\star}$ as follows
		\begin{equation}
		\EWb \approx
\frac{2\tL^{D-1}\rh}{\um^{d-1}}  \mathcal{S}_{\text{th}} \int_{\frac{r_d}{\rh}}^{\frac{r_w}{\rh}} \frac{ \dd{u}}{\Ab},\hspace{1cm}
		t-t_\star\approx  \frac{\Qb^{\star} \rh^z \um^{d+2z-3}}{g(\um)}\int_{\frac{r_d}{\rh}}^{\frac{r_w}{\rh}}\frac{\dd{u}}{{\Ab}}\;.
			\end{equation}
Now combining the above two equations
yields the following
\begin{equation}\label{eq:linear-Lif1}
   \Delta \EW\approx \tL^{D-1}   \mathcal{S}_{\text{th}}  \frac{2g(\um)}{\Qb^{\star} \um^{2d+2z-4}\rh^{z-1}}\; (t-t_{\star}).
\end{equation}
Expressing this formula in terms of the boundary quantities yields
\begin{equation}\label{eq:linear-Lif2}
\Delta \EW\approx \tL^{D-1}    \mathcal{S}_{\text{th}}  V_{W} \;(t-t_{\star}),
\end{equation}
where the velocity of the linear-growth depends on both $z$ and $\theta$ (via $d=D-\theta+1$) as follows
	\begin{equation}\label{vwz}
		V_{W}=  \qty(\frac{4\pi T}{d+z-1})^{\frac{z-1}{z}}\qty(\frac{d+z-3}{2(d+z-2)})^{\frac{d+z-2}{d+z-1}}\sqrt{\frac{d+z-1}{d+z-3}}.
	\end{equation}
 We note that for $z\neq 1$ the above velocity depends on the temperature of the final equilibrium state. 	Interestingly it is just the velocity of the linear-growth for entanglement entropy in the presence of Lifshitz and  hyperscaling-violating 
 exponents \cite{Fonda:2014ula,Alishahiha:2014cwa}. Indeed, in these references, the final result for HEE in the linear-growth regime was written in terms of the horizon radius (similar to eq. \eqref{eq:linear-Lif1}), thus the first factor in eq. \eqref{vwz} was neglected. 

\begin{figure}
	\begin{center}
		\includegraphics[scale=.74]{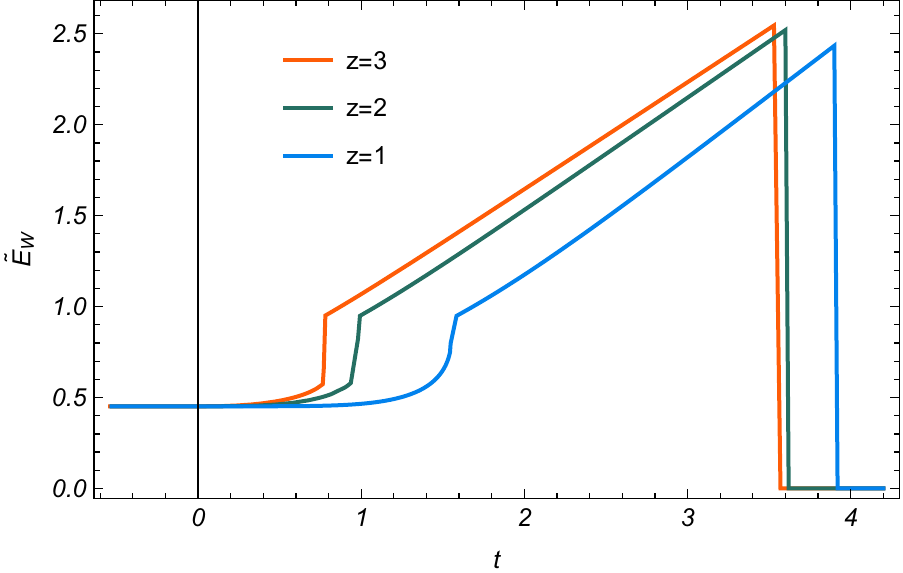}
%\hspace{0.2cm}
%  \includegraphics[scale=.74]{Vw.pdf}
			\end{center}
	\caption{Evolution of EWCS for different values of $z$. Here we set $h=2.1, \ell=4$ and $\rh=1$.}
	\label{fig:ewcsl4h21z1z2z3}
\end{figure}

\subsection*{Saturation}
As we explained in the previous section for the relativistic case, when the null shell reaches the turning point of $\Gamma_{2\ell+h}$, EWCS in the nonrelativistic background also saturates its equilibrium value. As far as the connected configuration is favored, the resultant value of $\EW$ is given by 
eq. \eqref{equilib}
while for a disconnected configuration, it vanishes abruptly. In figure 	\ref{fig:ewcsl4h21z1z2z3} we show the evolution of EWCS for different values of the dynamical exponent when $\Sigma$ becomes trivial at late times.

\section{Conclusions and Discussions}\label{sec:results}

In this work, we have studied the evolution of information measures dual to EWCS after a global quantum quench in holographic theories. In our
investigations, we have been focused on both relativistic boundary theories as well as non-relativistic ones which has nontrivial Lifshitz and hyperscaling-violating exponents. 
We present a combination of analytic and numerical results for symmetric strip-shaped
boundary subregions which enable us to study  different regimes of evolution during the thermalization process. 

In the limit of large entangling regions, we realize that the time evolution of EWCS in relativistic theories is characterized by three different scaling regimes: an early time quadratic growth, an intermediate linear-growth and a late time saturation. We found that as the width of the boundary region becomes larger, the region with linear-growth becomes more pronounced. Further, in theories with nontrivial dynamical and hyperscaling --violating exponents, the general behavior of the EWCS is very similar to the relativistic case, but in this case, the scaling of the initial growth is not quadratic and depends on $z$.  More explicitly, in this regime, we found $E_W \propto t^{1+\frac{1}{z}}$ which shows that the scaling of EWCS becomes less pronounced when the dynamical exponent became large. In particular, in $z\rightarrow \infty$ limit we have a linear-growth regime even in early times. On the other hand, during the intermediate stage of time evolution, EWCS exhibits a linear scaling whose coefficient depends on the thermal entropy density of the final
equilibrium state. Motivated by this linear-growth and in analogy to the previous results for HEE and the EWCS \cite{Liu:2013qca,BabaeiVelni:2020wfl} we define a rate of growth
\begin{equation}
\mathcal{R}_W(t)\equiv\frac{1}{\mathcal{S}_{\text{th}}\,l^{d-2}}\dv{E_W}{t}.
\end{equation}
Using eqs. \eqref{eq:early-Lif} and \eqref{eq:linear-Lif2} we find that for the HSL background
\begin{equation}
\mathcal{R}_W(t)=\Bigg\{ \begin{array}{rcl}
&\frac{2\pi}{d-1}\frac{\mathcal{E}}{\mathcal{S}_\text{th}}(zt)^{\frac{1}{z}}&\;\;\;t\ll \rh \ll \ell\\
&V_W&\;\;\;\rh \ll t\ll \ell
\end{array},
\end{equation}
where the entanglement velocity is given by eq. \eqref{vwz}. Specifically, We have shown that in nonrelativistic theories \textit{i.e.,} $z>1$ this velocity depends on the temperature of the final
equilibrium state. Indeed, we observe that $V_W\propto T^{\frac{z-1}{z}}$ and hence the larger the value of the temperature is, the faster EWCS grows in time. Moreover, in the asymptotic limit $z \rightarrow \infty$ the rate of growth has a linear dependence on temperature.  
An interesting question is if either
of these behaviors can be extracted from field theory calculations of various boundary information quantifiers dual to EWCS. Indeed, the time evolution of EE and some other related measures for nonrelativistic field theories with nontrivial Lifshitz exponent has been studied in \cite{MohammadiMozaffar:2018vmk,MohammadiMozaffar:2019gpn,Mozaffar:2021nex}. It would be interesting to figure out what the universal features of entanglement and information evolution in these theories are. 

\section*{Acknowledgments}
We would like to thank Ali Mollabashi for  his  collaboration in the early  stages  of  this  project.

\end{document}